# Rapport de recherche

## Un protocole de routage à basse consommation d'énergie pour les MANETs

01/10/2008

Saloua CHETTIBI

E-MAIL : atchettibi@gmail.com

Université Mentouri, Constantine, Algérie.

## ملخص

إن تمديد مدة خدمة الشبكات اللامركزية المتحركة يعد من أهم التحديات التي تطرح حين وضع تصور لبروتوكولات توصيل المعطيات، و هذا لأن الأجهزة النقالة تعمل بواسطة بطاريات محدودة الطاقة. كما أنه من الصعب غالبا تغيير أو إعادة شحن هذه الأخيرة خاصة حينما يتعلق الأمر بساحات المعارك أو المناطق المنكوبة، الخ.

إن استهلاك الطاقة لطالما كان يعد مكافئا لاستهلاك نطاق الموجات، لكن أبحاثا حديثة أثبتت أن "الطاقة" و" نطاق الموجات" وحدتان مختلفتان. بالإضافة إلى هذا، وجد أن الاستراتيجيات التقليدية المستعملة في توصيل المعطيات مثل التي تعتمد على "أقصر طريق" يمكن أن يكون لها تأثيرا سلبيا على توازن استهلاك الطاقة. من هذا المنطلق تم اعتماد مقاربات جديدة تتناول بصفة مباشرة مشكل الطاقة.

بحثنا هذا متعلق بإشكالية توصيل المعطيات مع فقد طاقة أقل. البروتوكول المقترح "MEA-DSR" يعتمد على إستراتيجية لتوزيع المهام بين الأجهزة النقالة بغرض تمديد مدة خدمة الشبكة. لتحقيق هذا استعملنا التوصيل المتعدد الطرق، بالإضافة إلى ذلك تم أخذ الطاقات المتبقية و طول الطرق بعين الاعتبار خلال اختيار طرق التوصيل. النتائج التي أسفرت عنها المحاكاة أبرزت فعالية البروتوكول المقترح وهذا تحت سيناريوهات صعبة تتميز بحركية وكثافة عاليتين و كذا تحت حجم اتصالات مرتفع.

**كلمات مفتاحية**: شبكات لامركزية متحركة، بروتوكولات التوصيل، استهلاك الطاقة.

## Résumé

L'allongement de la durée de vie du réseau constitue un grand défi dans la conception des protocoles de routage pour les réseaux mobiles ad-hoc (*MANETs*), car les noeuds mobiles sont équipés par des batteries dont la capacité est limitée. En outre, dans des environnements critiques (par exemple : champs des batailles, zones sinistrées, etc.) remplacer ou recharger les batteries est souvent impossible.

La consommation d'énergie a été considérée pour longtemps comme équivalente à la consommation de la bande passante. Cependant, des travaux plus récents ont montré que "énergie" et "bande passante" sont deux métriques différentes. De plus, il a été trouvé que les politiques de routage classiques comme celle du " plus court chemin" peuvent avoir un impact négatif sur l'équilibrage de la consommation d'énergie. Ainsi, plusieurs nouvelles approches de routage ont été proposées adressant explicitement la contrainte énergétique.

Notre travail est lié à la problématique de routage à basse consommation d'énergie pour les MANETs. Le protocole proposé MEA-DSR (*Multipath Energy-Aware on Demand Source Routing*) repose sur une politique de distribution de la charge entre les nœuds mobiles afin d'allonger la durée de vie du réseau. Pour atteindre cet objectif, on a fait recourt au routage multi-chemins ; les énergies résiduelles des nœuds mobiles ainsi que la longueur des chemins ont été également considérées lors de la prise des décisions de routage. Les résultats des simulations ont montré l'efficacité du protocole proposé dans des scénarios difficiles caractérisés par une haute mobilité, une haute densité et un trafic important.

**Mots-clés :** Réseaux Mobiles Ad-Hoc, Protocoles de Routage, Consommation d'Énergie.

## Abstract

Maximizing network lifetime is a very challenging issue in routing protocol design for Mobile Ad-hoc NETworks (*MANETs*), since mobile nodes are powered by limited-capacity batteries. Furthermore, replacing or recharging batteries is often impossible in critical environments (e.g. battlefields, disaster areas, etc.).

Energy consumption was considered for a long time equivalent to bandwidth consumption. However, recent works have shown that "energy" and "bandwidth" are substantially different metrics. Moreover, it was found that traditional routing policies such as "the shortest path" one can have a negative impact on energy consumption balance. Therefore, several new approaches have been proposed addressing energy efficiency explicitly.

Our work is related to energy efficient routing for MANETs' problem. The proposed MEA-DSR (*Multipath Energy-Aware on Demand Source Routing*) protocol is based on a load sharing strategy between mobile nodes in order to maximize network lifetime. To achieve this goal, we used multipath routing; nodes' residual energies and paths length were also considered when making routing decisions. Simulation results have shown the efficiency of the proposed protocol under difficult scenarios characterised by high mobility, high density and important traffic load.

**1 Introduction**

Le routage dans les MANETs s'effectue en mode multi-sauts ; des nœuds intermédiaires sont indispensables pour assurer la communication entre les nœuds sources et destinations qui ne résident pas dans la zone de transmission les un des autres. Dans un MANET, l'épuisement de l'énergie d'un nœud n'affecte pas uniquement sa capacité de communication mais peut carrément causer le partitionnement du réseau. L'objectif d'allongement de la durée de vie du réseau ne peut être achevé que si tous les nœuds mobiles sont traités équitablement.

Dans ce chapitre, nous présenterons le protocole de routage MEA-DSR [1,2] (*Multipath Energy-Aware on Demand Source Routing*) pour les MANETs. MEA-DSR est un protocole de routage unicast et réactif. Il est doté d'un mécanisme de distribution de la charge qui assure une exploitation plus équitable de l'énergie des nœuds mobiles. Ce mécanisme consiste à choisir les chemins composés de nœuds riches en énergie tout en exploitant la diversité des chemins qui peuvent exister entre une paire source-destination communicant.

La simulation du protocole MEA-DSR était conduite sous différents scénarios de mobilité, de densité et du trafic. De plus, ses performances étaient sujettes de comparaison avec celles du protocole DSR [3]. L'objectif étant d'optimiser la consommation d'énergie sans détérioration importante d'efficacité du réseau en termes du taux de délivrance des paquets de données et du délai bout-en-bout moyen.

**2 Motivations**

L'intuition derrière MEA-DSR est que la rupture des chemins dans les MANETs est une norme plutôt qu'une exception. Dans les protocoles de routage réactifs conventionnels comme DSR [3] et AODV [4], une grande quantité d'énergie est perdue à cause des fréquentes tentatives de découverte de chemins.

En outre, malgré que les changements fréquents de la topologie dans les MANETs permettent une sorte de distribution de la charge en forçant les nœuds sources à découvrir des nouveaux
chemins, mais certains nœuds sont sur-utilisés juste parce qu'ils possèdent des positions stratégiques dans le réseau.

Dans MEA-DSR, le routage multi-chemins est exploité pour minimiser la fréquence des découvertes de chemins ce qui assure un gain en consommation globale d'énergie. L'utilisation des chemins "maximalement-nœud-disjoints" fait partie de la politique de distribution de la charge adoptée dans MEA-DSR. Cette dernière, exploite également des informations sur les énergies résiduelles des nœuds et sur la longueur des chemins lors de la prise des décisions de routage. L'adoption d'un mécanisme de distribution de la charge permet une optimisation de la durée de vie de chaque nœud et par conséquence de la durée de vie de l'ensemble du réseau.

**3 Le protocole MEA-DSR**

Dans son fonctionnement de base, le protocole MEA-DSR est très similaire au protocole DSR. En réalité, DSR est un protocole de routage multi-chemins. Cependant, dans DSR plusieurs chemins sont stockés trivialement sans aucune contrainte de quantité ou de qualité. MEA-DSR limite le nombre de chemins que fournit un nœud destination à un nœud source à deux ; Il a été montré dans la référence [5] que l'utilisation de plus de deux chemins alternatifs ne présente qu'une amélioration de performances minime.

Le choix du premier chemin dans MEA-DSR est conditionné par deux facteurs : i) l'énergie résiduelle des nœuds constituant le chemin et ii) la puissance de transmission totale nécessaire pour transmettre des données sur ce chemin. Si on considère que les nœuds transmettent tous avec leurs puissances maximales (la propriété d'ajustement de la puissance de transmission n'est pas supportée par toutes les interfaces de communication sans fil), ce dernier facteur sera équivalent à celui du nombre de nœuds constituant le chemin. Quant au choix du deuxième chemin, le taux de disjonction par rapport au premier chemin vient en premier intérêt. Si plusieurs chemins présentent le même taux de disjonction, un sera sélectionné selon le même critère utilisé lors du choix du chemin primaire.

Au lieu de distribuer le trafic sur plusieurs chemins, dans MEA-DSR un seul chemin est utilisé à la fois pour l'acheminement des paquets de données. Cela permet d'éviter les problèmes du couplage des chemins, de congestion dans les nœuds communs (car il n'y a aucune garantie que le premier chemin et celui alternatif soient complètement disjoints) et de l'arrivée en désordre des paquets à leurs destinations.

### 3.1 Les paquets et les structures de données utilisés dans MEA-DSR

Les nœuds mobiles utilisant MEA-DSR échangent trois types de paquets de contrôle, à savoir : les requêtes de chemins (*RREQs*), les réponses de chemin (*RREPs*) et les erreurs de chemin (*RERRs*). Les mêmes formats des paquets RERR et RREP définis dans [6] pour le protocole DSR sont réutilisés dans MEA-DSR, tandis que le format des paquets RREQ a été légèrement modifié. Trois structures de données sont utilisées dans MEA-DSR : les caches de chemin, les tables des requêtes et les tables des chemins. Pour les caches aucune modification n'a était effectué sur le format défini dans [6]. Le format des tables des requêtes a été enrichi par des champs supplémentaires ; la table des chemins est une nouvelle structure spécifique à MEA-DSR.

#### 3.1.1 Paquet RREQ

Un champ nommé "*min_bat_lev*" a été ajouté au paquet RREQ ; il prend comme valeur le minimum des énergies résiduelles des nœuds parcourus par le paquet RREQ.

#### 3.1.2 Table des RREQs

Le format défini dans [6] a été augmenté par les champs : "*nb_hops*" dont la valeur représente le nombre de nœuds qu'a traversé le paquet *RREQ* et "*last_node"* qui maintient l'identificateur du voisin qui a transmis le paquet *RREQ*.

#### 3.1.3 Table des chemins

C'est une structure qui sert à stocker tous les chemins candidats au niveau des nœuds destinations, indexée par identificateur du noeud source. Chaque entrée dans cette table est constituée des champs suivants :

- *Src :* maintient l'identificateur du nœud source initiateur de la découverte de chemin.
- *Seq :* contient le numéro de séquence du RREQ.
- *Route :* la séquence des nœuds parcourus par le RREQ.
- *Min_bat_lev :* minimum des énergies résiduelles des nœuds parcourus par le paquet RREQ.
- *Arriving_time :* maintient le temps d'arrivée du RREQ.

Le contenu des quatre premiers champs est directement extrait de chaque RREQ reçu.

**3.2 Description du MEA-DSR**

Le protocole MEA-DSR est constitué de trois phases: i) la découverte des chemins, ii) la sélection des chemins par le nœud destination, et iii) la maintenance des chemins.

**3.2.1 Découverte des chemins**

La procédure de découverte des chemins dans MEA-DSR est similaire à celle du DSR. Si le nœud source a besoin d'un chemin vers une destination alors qu'aucun n'est déjà stocké dans sa cache, il broadcast un paquet RREQ à tous ses nœuds voisins. Dans MEA-DSR, seuls les nœuds destinations doivent répondre à un paquet RREQ car il sera difficile de contrôler la disjonction des chemins si les nœuds intermédiaires répondent de leurs caches comme c'est le cas dans DSR.

Pour éviter le problème de chevauchement des chemins [7], les nœuds intermédiaires ne suppriment pas toutes les copies des paquets RREQ comme dans DSR mais transfère ceux provenant sur un lien différent et dont le nombre de nœuds traversés soit inférieur ou égale à celui du premier paquet RREQ reçu. Cependant, transférer toutes les copies qui satisfont le critère expliqué ci-avant engendre un surdébit très important. De ce fait, nous avons limité le nombre des copies à transférer à un.

Quand un nœud intermédiaire situé au voisinage du nœud source reçoit un paquet RREQ, il insère la valeur de son énergie résiduelle dans le champ *"min_bat_lev"* du RREQ. Les autres nœuds intermédiaires qui reçoivent le RREQ, comparent leurs énergies résiduelles avec la valeur du champ *"min_bat_lev"*. Si c'est inférieur, ils substituent la valeur du champ "min_bat_lev" par leurs propres valeurs. Une fois la mise à jour du champ *"min_bat_lev"* soit achevée, le nœud intermédiaire ajoute son identificateur au RREQ et le re-broadcast à ses voisins. Le même procédé sera répété jusqu'à ce que le paquet RREQ atteigne sa destination finale.

**3.2.2 Sélection des chemins**

Après la réception du premier paquet RREQ, le nœud destination attends pour un certain temps "WT" (*Wait_Time*) avant de commencer la sélection des chemins. Une fois ce délai expire, la destination choisi comme chemin primaire "$che\min_i$" celui qui satisfait :

$$\frac{min\_bat\_lev_i}{route\_length_i} = \max_{j=1,n}\left(\frac{min\_bat\_lev_j}{route\_length_j}\right) \qquad (5.1)$$

Avec, *n* est le nombre de chemins stockés dans la table des chemins.

Après la sélection du chemin primaire, le nœud destination transmet immédiatement le chemin choisi dans un paquet RREP au nœud source. Ensuite, il choisit un deuxième chemin qui soit maximalement nœud-disjoint du premier. S'ils en existent plusieurs, celui qui satisfait l'équation (5.1) sera choisi et inclus dans un paquet RREP qui sera envoyé au nœud source.

**3.2.3 Maintenance des chemins**

Comme un lien dans un chemin peut être rompu à cause de la mobilité, des fluctuations du canal de communication ou à cause des pannes des noeuds, il est très important de rétablir les chemins brisés immédiatement. Dans MEA-DSR, si un nœud intermédiaire détecte la rupture d'un lien il transmet un paquet RERR au nœud source. Les nœuds recevant le paquet RERR retirent de leurs caches toutes les entrées qui utilisent ce lien. Si le noeud source ne possède aucun chemin valide dans sa cache, il initie une nouvelle procédure de découverte de chemins.

---

***Algorithm  Process _RREQ;***
*node_id: current node identifier; node_energy: current node residual energy; Time: RREQ arrival time.*
**Begin**
   ***if*** *(node_id<>RREQ.dest)* ***then***
            ***if*** *((node_id $\in$ RREQ.route)* ***or***
            *(RREQ.seq < request_table_seq(RREQ.seq, RREQ.src)* ***or***
            *( (RREQ.seq == request_table_seq(RREQ.seq, RREQ.src)* ***and*** *( (arrive_on_same_link )* ***or***
            *((arrive_on_different _link)* ***and*** *((route_ is_longer)* ***or***
            *(request_table_nb_copies(RREQ.seq,RREQ.src)==2)))* ***then***
                *Discard(RREQ);*
        ***else***
              *request_table_insert(RREQ.src, RREQ.route[RREQ.route.length()-1],*
                      *RREQ.seq,RREQ.route.length());*
            *Add (RREQ.route, node_id);*
             ***if*** *(RREQ .min_bat_lev > node_energy )* ***then***
                  *RREQ. min_bat_lev= node_energy;*
             ***endif***
            *Broadcast(RREQ);*
        ***endif***

   ***else***
            *route_table_insert (RREQ.src,RREQ.seq,*
                    *RREQ.route,RREQ.bat_lev, Time);*
        *Discard(RREQ);*

        ***if*** *(wait_time is exhausted )* ***then***
            *Route1 $\leftarrow$ MAX_bat_length {route, route $\in$ route_table};*
            *Add (Route1, node_id);*
            *RREP1.route $\leftarrow$ Route1;*
            *Unicast (RREP1);*
            *ENS_Route2 $\leftarrow$ Max_disjoint_route1{route,route $\in$ route_table};*
            *Route2 $\leftarrow$ MAX_ bat_length{route, route $\in$ ENS_Route2 };*
            *Add (Route2, node_id);*
            *RREP2.route $\leftarrow$ Route2;*
            *Unicast (RREP2);*
        ***endif***
   ***endif***
**End**

---

**Figure 5.1.** Algorithme du traitement d'un paquet RREQ.

```
Algorithm  Process _RREP;
node_id: current node identifier;
Begin
      if (node_id<>RREP.dest) then
              Forward (RREP);
      else
              route_cache_insert (RREP.src, RREP.route);
              Discard(RREP);
      endif
End
```

**Figure 5.2.** Algorithme du traitement d'un paquet RREP.

```
Algorithm  Process _RERR;
node_id: current node identifier;
Begin
       update_current_node_cache;
       if (node_id<>RERR.dest) then
              Forward(RERR);
      else
              Discard (RERR);
       endif
End
```

**Figure 5.3.** Algorithme du traitement d'un paquet RERR.

**4 Évaluation des performances par simulation**

Dans les sections suivantes, nous commencerons par la description de l'ensemble des paramètres de simulation ainsi que les différentes métriques de performances auxquelles nous nous sommes intéressés. Ensuite, nous présenterons l'ensemble des résultats auxquels nous avons abouti tout en fournissant les interprétations nécessaires. Les performances du protocole MEA-DSR étaient sujettes de comparaison avec les performances du protocole DSR, qui est un parmi les protocoles de routage les plus acceptés dans le domaine. Ces comparaisons ont été menées sous des conditions différentes de mobilité de nœuds, de densité du réseau et du trafic de données .

**4.1 Environnement de simulation**

Pour l'évaluation des performances des protocoles étudiés, nous avons utilisé le simulateur NS version 2.32 [8] (une description du simulateur NS-2 est fournie en annexe A). Le réseau considéré est composé de 50 nœuds mobiles (ce paramètre sera varié dans la deuxième série d'expérimentations) déployés sur une surface de 1000mx1000m. Nous avons supposé que tous les nœuds mobiles du réseau sont équipés par des interfaces de communication IEEE 802.11 avec un débit de 2MBps utilisant la DCF (*Distributed coordination function*) de MAC avec RTS et CTS, et que chaque nœud possède une zone de transmission de 250 m. Comme modèle de propagation d'ondes, nous avons utilisé le "Two Ray Ground model " déjà décrit dans le paragraphe (§ 2.7.8.2) (le script qui décrit la configuration du réseau et qui analyse les résultats de la simulation est fourni en annexe B).

Le modèle de mobilité utilisé dans toutes les simulations était le RWP déjà décrit dans le paragraphe (§ 2.7.8.1). Chaque nœud se déplace avec une vitesse qui varie uniformément dans l'intervalle [5 m/s, 10m/s] avec une durée de pause de 100s (ces paramètres seront variés dans la première série d'expérimentations). La durée de chaque simulation était de 600 s. Chaque point sur un graphe correspond à la moyenne de cinq exécutions avec des scénarios de mobilité différents générés aléatoirement.

Nous avons choisi de modéliser la communication entre les nœuds en utilisant le trafic CBR sur UDP, où chaque source génère des paquets de 512 octets avec un taux de 4 paquets par seconde (ce taux sera varié dans la troisième série d'expérimentations). Un total de 10 connexions a été généré (ce nombre sera également varié dans la troisième série d'expérimentations); ces connexions commencent dans un temps aléatoirement choisi de l'intervalle [0s ,120s] et restent actives jusqu'à la fin de la simulation (le script correspondant est fourni en annexe C). Notons que TCP n'a pas été utilisé car il modifie le temps d'émission des paquets selon sa perception de la capacité du réseau, ce qui empêche une comparaison directe des protocoles de routage en question.

Comme notre protocole n'adresse pas le problème d'énergie consommée en mode inactif, nous n'avons considéré que l'énergie consommée dans les modes de réception et d'émission. Comme valeurs, nous avons utilisé ceux obtenues via des expériences dans un travail antérieur [9] à savoir : 1W pour le mode réception, et 1.4W pour le mode transmission. De plus Nous avons considéré que l'énergie initiale de chaque nœud est de 1000J. Notons que la valeur du WT dans le protocole MEA-DSR a été fixée expérimentalement à 0.06 s (l'annexe D donne une idée sur l'influence du paramètre WT sur les performances du protocole MEA-DSR).

## 4.2 Les métriques de performance

Nous nous sommes intéressés à mesurer les métriques de performances suivantes :

- *Le Surdébit de Routage Normalisé (SRN) :* c'est le ratio entre le nombre de paquets du contrôle (*RREQs, RREPs, RERRs*) générés par le protocole de routage et le nombre de paquets de données bien reçus.
- *Le Taux de Délivrance (TD)* : c'est le rapport entre le nombre des paquets de données bien reçus par les nœuds destinations à celui généré par les nœuds sources.
- *Le Délai bout-en-bout Moyen (DM)* : c'est la moyenne des différences entre le temps d'arrivée d'un paquet de donnée à sa destination et le temps de son émission par le nœud source, pour tous les paquets de données bien reçus dans le réseau.
- *L'Énergie Consommée par Paquet (ECP) :* représente le rapport entre l'énergie totale consommée dans le réseau et le nombre de paquets de données bien reçus.
- *L'Écart Type d'Energie Consommée par Nœud (ETECN) :* c'est la racine carrée de la moyenne des carrées des différences entre l'énergie consommée par chaque nœud et la moyenne des énergies consommée par nœud. Cette métrique est utilisée pour évaluer l'équité d'utilisation des nœuds par le protocole de routage en question. Plus l'écart type est petit, plus est équitable la consommation d'énergie.
- *Le Taux d'Énergie Résiduelle Minimal (TERM) :* il s'agit du minimum des taux des énergies résiduelles des nœuds par rapport à leurs énergies initiales à la fin de la simulation.

## 4.3 Étude de l'impact de la mobilité des nœuds sur les performances des protocoles MEA-DSR et DSR

Dans cette première série d'expérimentations, nous nous intéressons à l'étude de l'influence de la mobilité des nœuds sur les performances des protocoles MEA-DSR et DSR. Dans cette perspective, nous avons varié la durée de la pause entre 0 s (mobilité continue) jusqu'à 600 s (aucune mobilité). De plus, nous avons mené des simulations pour les cas d'une:

- Faible vitesse : varie uniformément dans l'intervalle [0,5 m/s, 1 m/s].
- Vitesse moyenne : varie uniformément dans l'intervalle [5 m/s, 10 m/s].
- Haute vitesse : varie uniformément dans l'intervalle [20 m/s, 25 m/s].

Dans tout ce qui suit, nous désignons par forte mobilité (faible mobilité) : une haute vitesse (une faible vitesse) et/ou une mobilité fréquente (mobilité peu fréquente) qui est fonction de la durée de la pause. Augmenter la vitesse ou diminuer le temps de la pause provoquent (diminuer la vitesse ou augmenter le temps de la pause), tous les deux, des changements fréquents (peu fréquents) dans la topologie du réseau.

### 4.3.1 Surdébit de routage normalisé

Sur la fig.5.4, nous constatons que DSR génère plus de surdébit de routage sous des conditions d'une forte mobilité par rapport à MEA-DSR. Cela est dû au nombre important des paquets RREQ*s* que génère DSR en réinitialisant à chaque fois des découvertes de chemins.

Dans des scénarios d'une faible mobilité, les chemins deviennent de plus en plus robustes ce qui diminue le besoin d'initialisation des découvertes de chemins pour les deux protocoles. Cependant, MEA-DSR continu à générer un surdébit important. Cela parce que les nœuds intermédiaires dans MEA-DSR transmettent des copies des paquets RREQs, tandis que dans DSR toutes les copies sont supprimées. De plus, les paquets RREQs dans MEA-DSR se propagent toujours jusqu'à leurs destinations alors que dans DSR les nœuds intermédiaires peuvent directement répondre à partir de leurs caches.

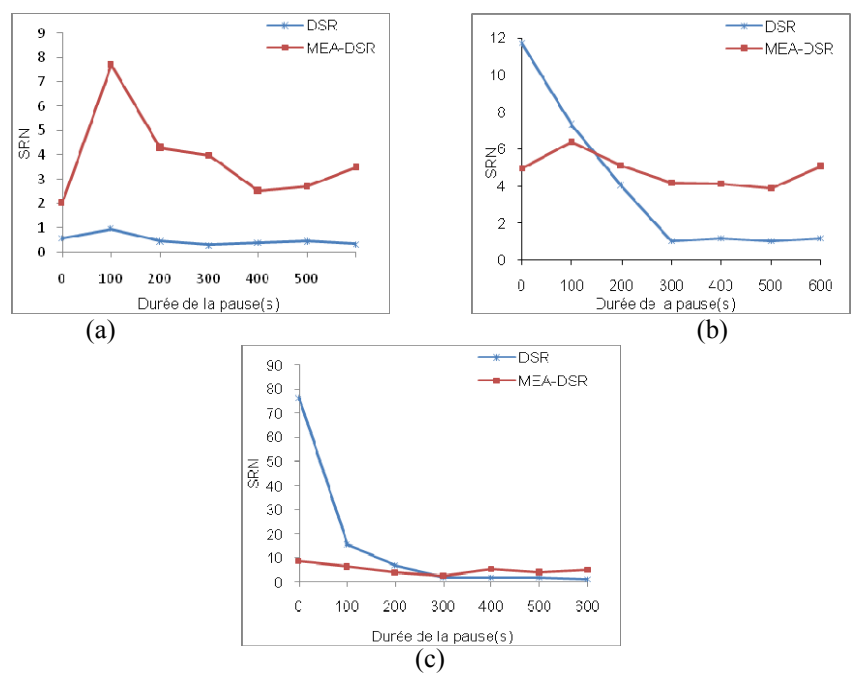

**Figure 5.4.** Surdébit de routage normalisé vs durée de la pause :
(a) cas d'une faible vitesse (b) cas d'une vitesse moyenne (c) cas d'une haute vitesse.

**4.3.2 Taux de délivrance**

Sur la fig.5.5, on observe que pour des scénarios d'une forte mobilité, MEA-DSR offre un TD plus important que celui du DSR. Cela parce que dans DSR les nœuds intermédiaires sont autorisés à répondre par des chemins stockés dans leurs caches qui sont, malheureusement, souvent invalides. Par conséquence, les paquets de données émis sur ces chemins seront supprimés dès qu'ils atteignent les liens brisés, du moment que même le mécanisme du salvaging devient inefficace. De plus, les paquets de données dans DSR subissent des délais supplémentaires au niveau des files d'attente des interfaces de communication à cause des retransmissions fréquentes et des tentatives du salvaging répétitives. Cette latence cause l'expiration des paquets (i.e. leur suppression).

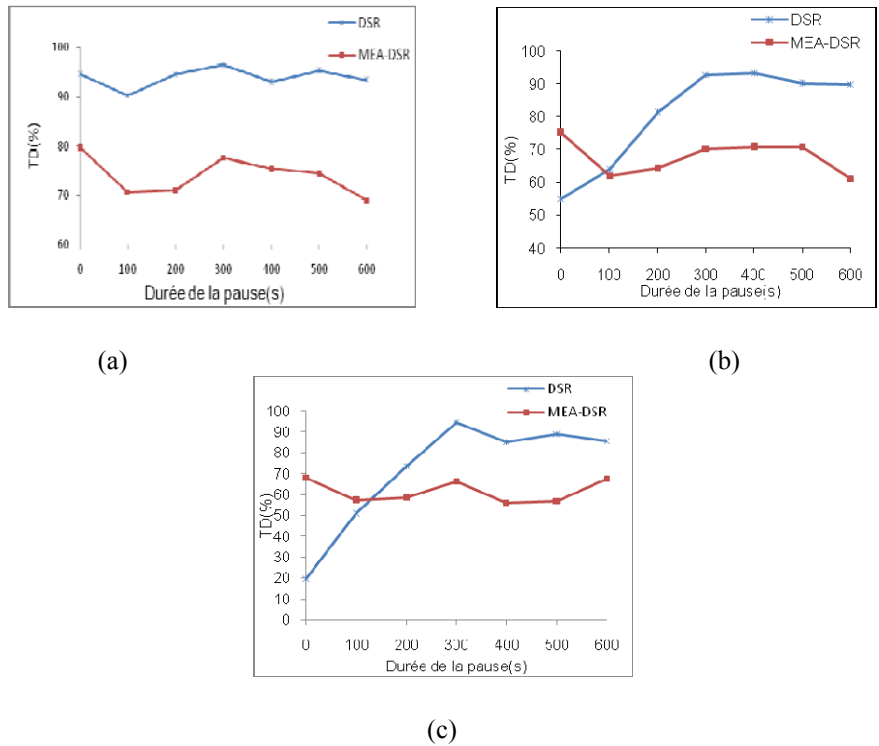

(a)  (b)

(c)

**Figure 5.5** Taux de délivrance vs durée de la pause :
(a) cas d'une faible vitesse (b) cas d'une vitesse moyenne (c) cas d'une haute vitesse.

Pour des scénarios de mobilité plus basse, le TD du DSR s'augmente pour qu'il dépasse celui du MEA-DSR car plus le réseau est stable, plus les chemins sont robustes et plus deviennent efficaces les mécanismes du salvaging et de réponse à partir des nœuds intermédiaires du DSR. Dans MEA-DSR, les nœuds intermédiaires ne sont pas autorisés d'utiliser leurs caches pour retransmettre les paquets, de ce fait la probabilité de suppression des paquets reste supérieure à celle dans DSR.

**4.3.3 Délai bout en bout moyen**

Sur la fig.5.6, on observe que dans le cas d'une haute mobilité, le DM du DSR est le plus élevé. Cela est dû aux délais supplémentaires que subissent les paquets de données au niveau des files d'attente des interfaces de communication à cause des retransmissions fréquentes et des tentatives du salvaging répétitives. De plus, les paquets de données dans DSR passent plus de temps dans les files d'attente des nœuds sources en l'attente d'établissement des chemins.

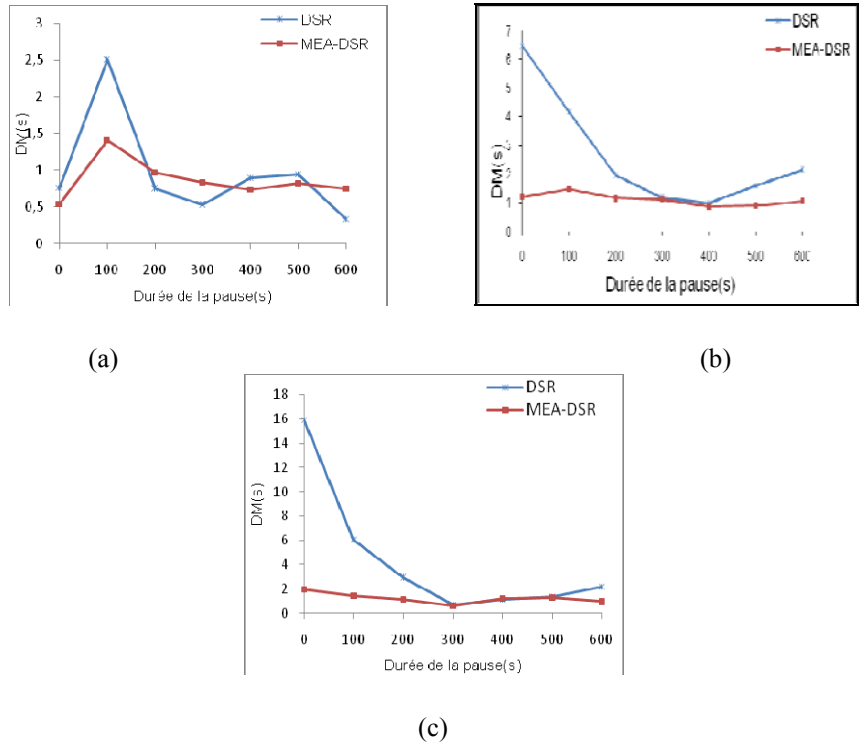

**Figure 5.6.** Délai bout-en-bout moyen vs durée de la pause :
(a) cas d'une faible vitesse (b) cas d'une vitesse moyenne (c) cas d'une haute vitesse.

Dans des conditions d'une faible mobilité, le DM du DSR s'approche de celui du MEA-DSR car les reconstructions de chemins deviennent moins fréquentes.

**4.3.4 Énergie consommée par paquet**

L'ECP donne une idée sur la consommation globale d'énergie dans le réseau. Elle est proportionnelle au surdébit de routage généré par chaque protocole et à la longueur des chemins utilisés. Pour des scénarios d'une forte mobilité, DSR génère plus de surdébit que MEA-DSR ce qui fait qu'il consomme globalement plus d'énergie.

Pour des scénarios d'une mobilité plus basse, bien que DSR génère moins de surdébit par rapport à MEA-DSR mais il n'a pas marqué une amélioration importante en consommation d'énergie. Cela parce que DSR tend à utiliser des chemins plus longs qui viennent dans les RREPs générés par les nœuds intermédiaires (ces derniers effectuent des concaténations). De ce fait, la puissance de transmission totale d'un paquet reste élevée.

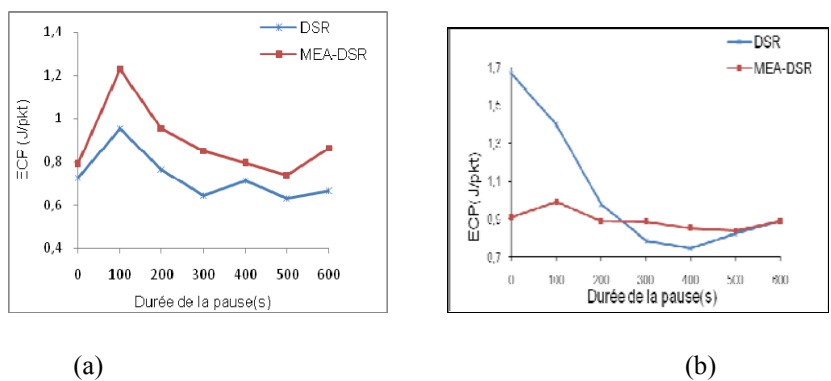

(a)                 (b)

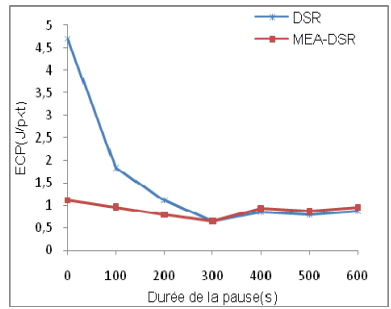

(c)

**Figure 5.7.** Énergie consommée par paquet vs durée de la pause :
(a) cas d'une faible vitesse (b) cas d'une vitesse moyenne (c) cas d'une haute vitesse.

### 4.3.5 Écart type d'énergie consommée par nœud

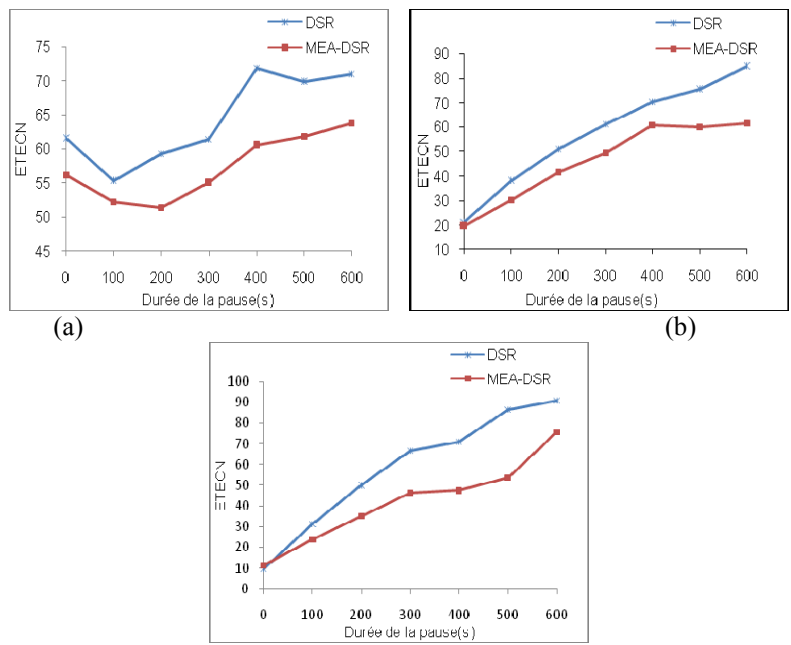

**Figure 5.8.** Ecart type d'énergie consommée par nœud vs durée de la pause :
(a) cas d'une faible vitesse (b) cas d'une vitesse moyenne (c) cas d'une haute vitesse.

Sur la fig.5.8, nous constatons que dans tous les scénarios de mobilité, l'écart type des énergies consommées dans MEA-DSR est inférieur à celui du DSR. Cela confirme l'efficacité du mécanisme de distribution de la charge (choix des chemins constitués de nœuds riches en énergie et utilisation de chemins les plus disjoint) utilisé dans MEA-DSR.

Pour les deux protocoles, plus le réseau est stable plus l'ETECN augmente. Cela était attendu, car un même chemin reste en utilisation tant qu'il n'est pas rompu.

### 4.3.6 Taux d'énergie résiduelle minimal

Le TERM donne une idée sur la tendance du protocole à maximiser la durée de vie des nœuds et donc de l'ensemble du réseau. L'équilibrage de la distribution de la charge qu'a présenté MEA-DSR lui a permis de marquer un TERM plus haut que celui du DSR, dans tous les scénarios de mobilité. Pour le cas d'une haute mobilité, le gain en énergie résiduelle est plus important grâce à la minimisation de la consommation d'énergie globale.

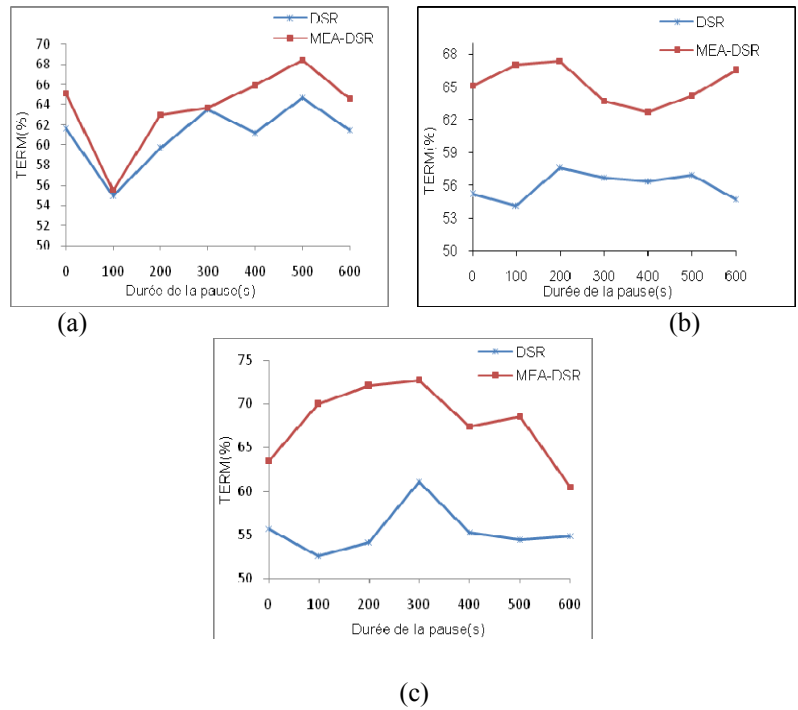

**Figure 5.9.** Taux d'énergie résiduelle minimale vs durée de la pause :
(a) cas de mobilité faible (b) cas de mobilité moyenne (c) cas de mobilité forte.

### 4.4 Étude de l'impact de la densité du réseau sur les performances des protocoles MEA-DSR et DSR

Nous avons mené un deuxième ensemble d'expérimentations en variant la densité du réseau à partir de 50 nœuds jusqu'à 100 nœuds.

#### 4.4.1 Surdébit de routage normalisé

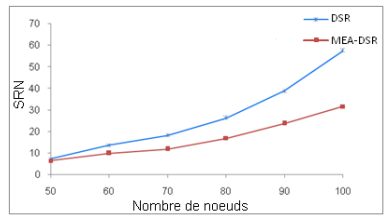

**Figure 5.10.** Surdébit de routage normalisé vs nombre de nœuds

Comme c'était attendu, le surdébit de routage des deux protocoles a augmenté avec l'augmentation de la densité du réseau. Il est claire sur la fig. 5.10 que quelque soit la densité, le SRN du MEA-DSR était inférieur que celui du DSR. Cela parce que MEA-DSR réinitie moins fréquemment des découvertes de chemins grâce à l'utilisation de chemins maximalement nœud-disjoints.

**4.4.2 Taux de délivrance**

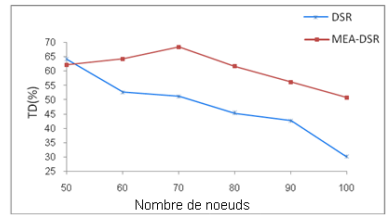

**Figure 5.11.** Taux de délivrance vs nombre de nœuds

Pour toutes les conditions de densité (à l'exception du cas de 50 nœuds où les TD des deux protocoles étaient presque égaux), le TD de MEA-DSR était supérieur que celui du DSR. Cela parce que DSR souffre plus que MEA-DSR de la congestion des files d'attente des interfaces de communication car il génère plus de surdébit (voir fig. 5.10). Le surdébit de routage des deux protocoles augmente en fonction du nombre de nœuds, chose qui a aggravé le problème de congestion des files d'attente. Ceci explique la réduction de TD des deux protocoles en augmentant la densité du réseau.

En examinant les fichiers traces des simulations, nous avons trouvé que MEA-DSR tend à utiliser des chemins plus long en augmentant la densité du réseau. Cela augmente la probabilité de rupture de chemins, et décrémente par conséquent son TD.

**4.4.3 Délai bout-en-bout moyen**

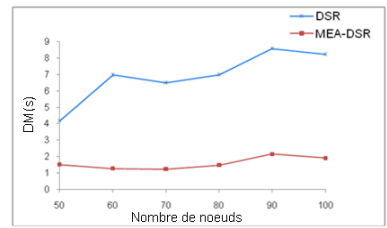

**Figure 5.12.** Délai bout-en-bout moyen vs nombre de nœuds

L'augmentation de la densité du réseau emmène à l'occurrence de plus de collisions. Par conséquent, dans les deux protocoles les paquets de données passent plus de temps dans les files d'attente des interfaces de communication à cause des fréquentes retransmissions.

Dans tous les cas, le DM du MEA-DSR était inférieur que celui du DSR. Cela peut être trivialement expliqué par le fait que les paquets de données dans DSR passent plus de temps dans les files d'attente des nœuds sources en l'attente d'établissement des chemins.

**4.4.4 Énergie Consommée par paquet**

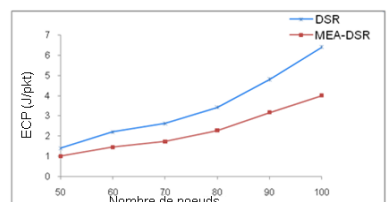

**Figure 5.13.** Énergie consommée par paquet vs nombre de nœuds

L'augmentation de la densité du réseau augmente le risque des interférences (de ce fait plusieurs retransmissions sont nécessaires) et cause une élévation du nombre des paquets de contrôle échangés dans le réseau. Tous ces facteurs justifient l'accroissement de l'ECP des deux protocoles.

Pour toutes les conditions de densité, l'ECP de MEA-DSR était inferieur que celui de DSR car MEA-DSR génère toujours moins de surdébit que DSR. Par conséquence, la consommation d'énergie globale dans MEA-DSR reste moins élevée.

### 4.4.5 Écart type de l'énergie consommée par nœud

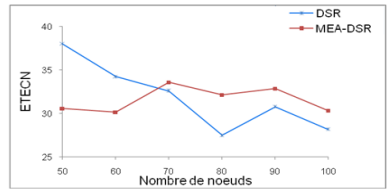

**Figure 5.14.** Écart type de l'énergie consommée par nœud vs nombre de nœuds

L'efficacité de la politique de distribution de la charge implémentée dans MEA-DSR est claire dans les cas de 50 et 60 nœuds, où l'ETECN de ce dernier était inferieur que celui du DSR. A partir du cas de 60 nœuds, l'ETECN de DSR s'améliorait grâce à la variété des RREPs arrivant des nœuds intermédiaires (l'augmentation de la densité augmente le nombre de voisins par nœud et donc engendre une variété dans les RREPs). Pour MEA-DSR, son ETECN a légèrement augmenté (en réalité son ETECN a varié entre les valeurs 30 et 33). Cela est dû à l'augmentation des interférences qui causent la perte des RREQs, chose qui diminue le nombre des chemins alternatifs parmi lesquels les nœuds destinations choisissent leurs réponses.

### 4.4.6 Taux d'énergie résiduelle minimal

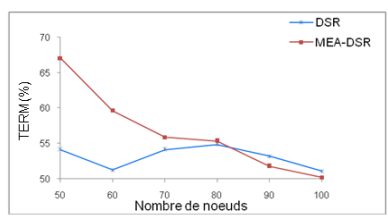

**Figure 5.15.** Taux d'énergie résiduelle minimal vs nombre de nœuds

TERM est fonction de la consommation globale d'énergie et de l'équité de distribution de la charge dans le réseau par le protocole de routage. Comme l'ETECN de MEA-DSR était presque le même dans tous les cas, la réduction du son TERM est principalement due à l'augmentation du son surdébit de routage.

Malgré que le surdébit de routage du DSR a augmenté également, mais ça n'a pas causé une réduction considérable de son TERM (il variait entre 51% et 54%). Ce résultat est justifié par l'amélioration de l'équité d'usage des nœuds par DSR. Néanmoins, le TERM du DSR était presque toujours inferieur que celui du MEA-DSR car il génère typiquement plus de surdébit que ce dernier.

**4.5 Étude de l'impact du trafic sur les performances des protocoles MEA-DSR et DSR**

Dans cette série d'expérimentations, nous étudions l'impact du trafic sur les performances des protocoles MEA-DSR et DSR. Dans cet objectif nous avons varié, premièrement, le taux d'envoi des paquets par seconde où le nombre des sessions de communication a été fixé à 10. Deuxièmement, nous avons fixé le taux d'envoi des paquets par seconde à 4 tandis que nous avons varié le nombre des sessions de communication.

**4.5.1 Surdébit de routage normalisé**

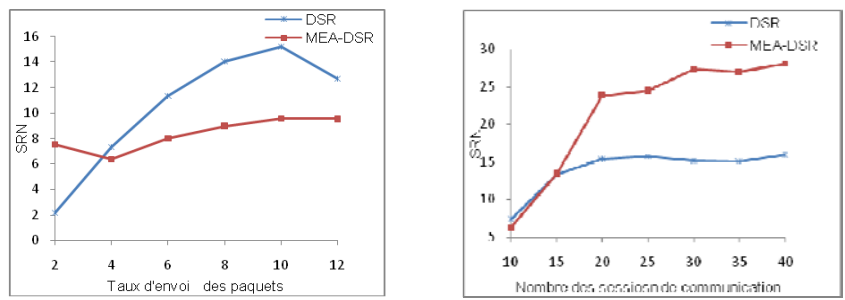

(b)                                                                                               (b)
**Figure 5.16.** Surdébit de routage normalisé vs :
(a) taux d'envoi des paquets (b) nombre des sessions de communication.

Si on augmente le taux d'envoi des paquets, il y aura plus de perte de paquets. Ces pertes seront interprétées comme des ruptures de chemins ce qui implique plus de reconstructions. Si on augmente le nombre des sessions de communication plus d'opérations par protocole seront initialisées. Dans les deux cas, le surdébit de routage augmentera. C'est ce que confirme la fig.5.16 pour les deux protocoles.

En augmentant le taux d'envoi des paquets, le SRN de MEA-DSR était inférieur que celui du DSR. Cela est dû au fait que DSR réinitie plus fréquemment des découvertes de chemins par session de communication. En augmentant le nombre des sessions de communication, le SRN du MEA-DSR était plus important que celui du DSR. Cela est normal, car MEA-DSR génèrent plus de surdébit par cycle de découverte de chemins en renvoyant des RREQs dupliqués.

**4.5.2 Taux de délivrance**

L'augmentation du trafic provoque des interférences et de la congestion, ce qui cause plus de perte des paquets. Cela explique pourquoi le TD des deux protocoles a diminué avec l'augmentation du taux d'envoi des paquets et du nombre des sessions de communication.

En augmentant le taux d'envoi des paquets, MEA-DSR surperforme DSR. Cela parce que MEA-DSR souffre moins du problème de la congestion des files d'attentes du moment qu'il génère moins de surdébit de routage par session de communication (voir fig.5.16.a)

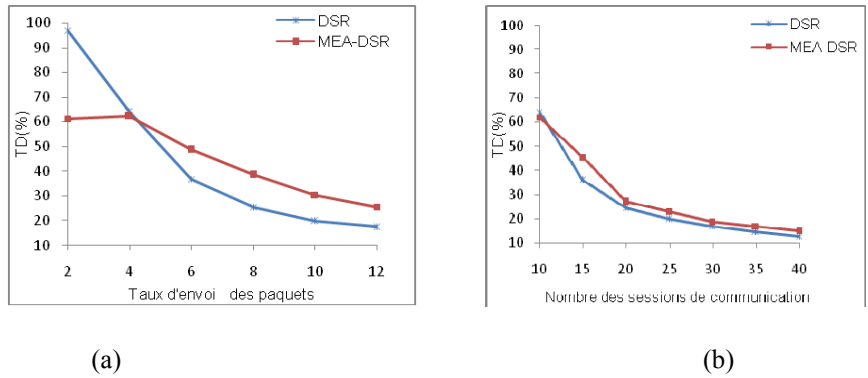

(a)                                                                         (b)

**Figure 5.17.** Taux de délivrance vs :
(a) taux d'envoi des paquets (b) nombre des sessions de communication.

Malgré que l'accroissement du nombre des sessions de communication a provoqué l'augmentation du surdébit généré par MEA-DSR (voir fig.5.16.b), mais cela n'a pas aidé DSR pour surperformer MEA-DSR. Le protocole DSR continue à supprimer plus de paquets de données à cause de la congestion des files d'attente (cela a été confirmé en examinant les fichiers traces des simulations). Ce comportement peut être expliqué par le fait que DSR passe plus de temps avant de libérer une entrée correspondante à des paquets de données non encore acquittés en effectuant plusieurs tentatives du salvaging (nous rappelons ici que en cas d'une haute mobilité les chemins cachés sont plus susceptibles d'êtres invalides). Durant ce moment, les nouveaux paquets de données arrivant vont être supprimés. Cela pose problème particulièrement quand les nœuds émetteurs de ces paquets ne maintiennent plus de chemins alternatifs valides.

**4.5.3 Délai bout-en-bout moyen**

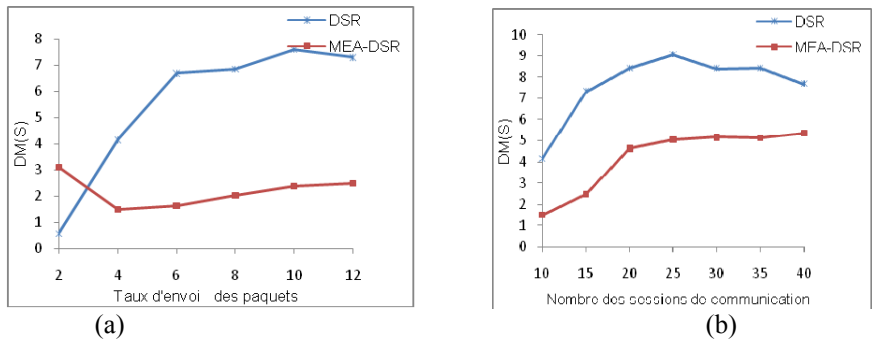

(a)                                                                         (b)

**Figure 5.18.** Délai bout-en-bout moyen vs :
(a) taux d'envoi des paquets (b) nombre des sessions de communication.

L'augmentation du trafic emmène à l'occurrence de plus de collisions et à plus de congestion. Par conséquence, les paquets de données passent plus de temps dans les files d'attente des interfaces de communication à cause des fréquentes retransmissions. Cela explique l'augmentation du DM des deux protocoles en augmentant le taux d'envoi des paquets et le nombre des sessions de communication.

Dans tous les cas, le DM du MEA-DSR est le meilleur. Cela peut être trivialement expliqué par le fait que les paquets de données dans DSR passent plus de temps dans les files d'attente des nœuds sources en l'attente d'établissement des chemins.

**4.5.4 Énergie Consommée par paquet**

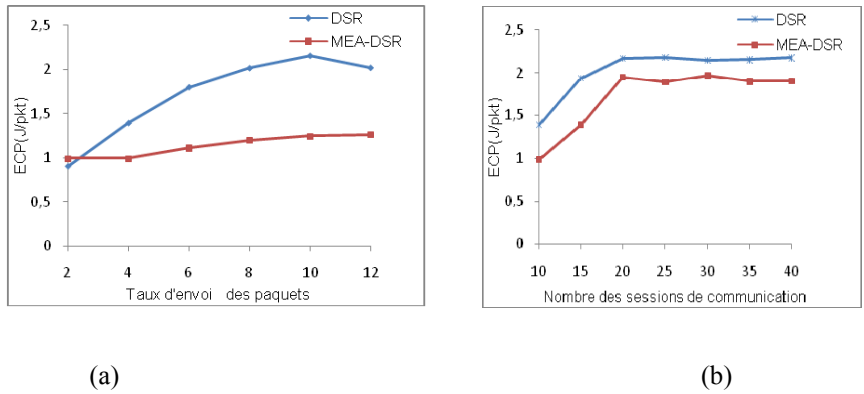

(a)                                                                 (b)

**Figure 5.19.** Énergie Consommée par paquet vs:
(a) taux d'envoi des paquets nombre de nœuds (b) nombre des sessions de communication.

En augmentant le taux d'envoi des paquets, l'ECP du MEA-DSR était la meilleure car il génère moins de surdébit de routage. Malgré qu'en augmentant le nombre des sessions de communication le surdébit généré par MEA-DSR était plus important que celui du DSR, mais cela n'as pas causé la surperformance du DSR. Cela est dû au fait que DSR consomme plus d'énergie en effectuant des tentatives du salvaging inefficaces.

**4.4.5 Écart type de l'énergie consommée par nœud**

L'efficacité de la politique de distribution de la charge implémentée dans MEA-DSR, en comparaison avec DSR, était limitée aux cas d'un faible trafic (taux d'envoi de 2pkt/s et 4pkt/s avec 10 connections). Dans tous les autres cas (taux d'envoi des paquets entre 8 pkts/s et 12pkts/s et nombre des sessions entre 15 et 40), l'ETECN du MEA-DSR était le pire. Cela parce que l'augmentation du trafic provoque des interférences et de la congestion. Chose qui cause plus de perte en paquets RREQs, ce qui diminue le nombre des chemins alternatifs parmi lesquels les nœuds destinations choisissent leurs réponses. En incrémentant le taux d'envoi des paquets, l'ETECN des deux protocoles a augmenté car plus du trafic est injecté entre les mêmes paires source-destination. De ce fait, les mêmes nœuds restent en utilisation pour longtemps.

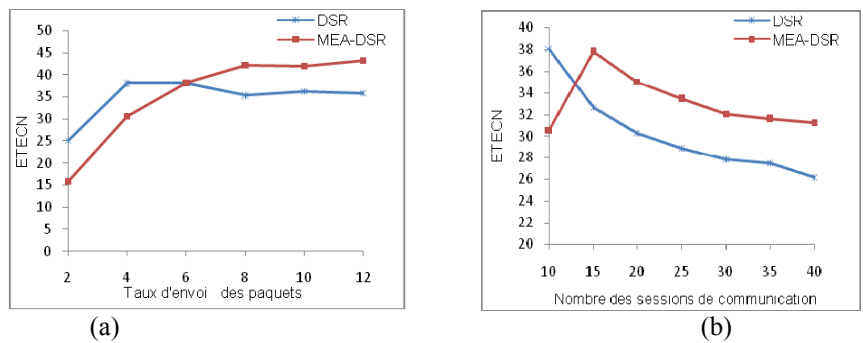

(a)                                                                 (b)

**Figure 5.20.** Écart type de l'énergie consommée par nœud vs *:*
(a) taux d'envoi des paquets (b) nombre des sessions de communication.

En incrémentant le nombre des sessions de communication, les deux protocoles ont marqué une amélioration dans leurs ETECN. Cela parce que plus de nœuds initialisent des communications, ce qui a donné des profiles de consommation d'énergie semblables.

### 4.4.6 Taux d'énergie résiduelle minimal

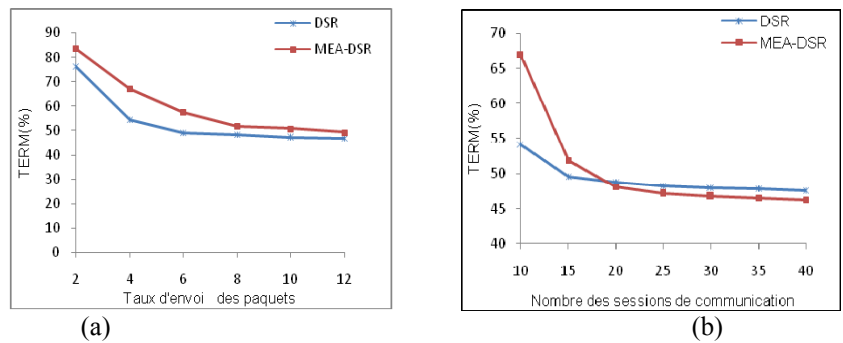

**Figure 5.21.** Taux d'énergie résiduelle minimal vs :
(a) taux d'envoi des paquets (b) nombre des sessions de communication.

TERM est fonction de l'énergie globale consommée et de l'équité en utilisation des nœuds par les protocoles de routage. Du moment que la consommation d'énergie globale du MEA-DSR était moins importante que celle du DSR (voir fig.5.19.a), le TERM du MEA-DSR était le plus élevé pour tous les taux d'envoi des paquets. Cela s'applique aussi aux cas du 10 et 15 sessions. A partir du point de 20 sessions, TERM du DSR s'est un peu amélioré grâce à l'amélioration de son équité en utilisation des nœuds.

## 5 Conclusion

Les chemins maximalement nœud-disjoints sont exploités dans MEA-DSR pour réaliser deux objectifs: i) minimiser la consommation d'énergie globale en réduisant la fréquence des découvertes de chemins, et par conséquence le surdébit de routage qui en découle ; et ii) équilibrer la consommation d'énergie entre les nœuds dans le réseau. La politique de distribution de la charge implémentée dans MEA-DSR consiste à munir les nœuds sources par deux chemins. Le chemin primaire doit avoir le meilleur rapport entre le minimum des énergies résiduelles des nœuds qui le composent et la longueur du chemin. Le chemin alternatif doit être maximalement nœud-disjoint du premier. Si plusieurs chemins présentent le même taux de disjonction, un parmi ces chemins qui satisfait le critère appliqué lors du choix du chemin primaire sera sélectionné.

Dans ce chapitre, nous avons présenté une description détaillée du protocole MEA-DSR. Dans les simulations nous nous sommes intéressés à l'évaluation des métriques de performances suivantes:

- ➢ Le surdébit de routage normalisé.
- ➢ Le taux de délivrance.
- ➢ Le délai bout-en-bout moyen.
- ➢ L'énergie consommée par paquet.
- ➢ L'écart type d'énergie consommée par nœud.
- ➢ Le taux d'énergie résiduelle minimal.

Les simulations ont été menées sous des conditions différentes de mobilité, de densité et du trafic, et les performances du protocole MEA-DSR ont été confrontées à celles du protocole DSR.

Les résultats auxquels nous avons abouti peuvent être résumés comme suit :

i) Impact de la mobilité des nœuds

Sous des conditions d'une forte mobilité, le protocole MEA-DSR était le plus performant pour toutes les métriques mesurées. Pour des conditions d'une faible mobilité, MEA-DSR a été énergétiquement plus efficace [1] que DSR mais au prix d'un surdébit de routage élevé. De plus, MEA-DSR a marqué un TD plus bas. Néanmoins, cela ne constitue pas une limitation pratique car dans des applications réelles les nœuds dans un MANETs sont souvent caractérisés par une forte mobilité.

ii) Impact de la densité du réseau

En augmentant la densité du réseau, les deux protocoles ont marqué une dégradation de leurs performances. MEA-DSR était le meilleur dans toutes les mesures, sauf pour l'ETECN où le protocole DSR a surperformé MEA-DSR à partir du point de 70 nœuds. Cela lui a permis de marquer un TERM plus élevé pour les cas de 90 et 100 nœuds. Cependant, la différence dans le TERM des deux protocoles n'était pas très significative.

iii) Impact du Trafic

En augmentant le trafic, les deux protocoles ont montré une dégradation de leurs performances. A partir d'un taux d'envoi supérieur à 2pkt/s avec un nombre de sessions de 10, MEA-DSR a marqué un SRN et un DM plus bas. Pour le TD, MEA-DSR a surperformé DSR à partir du point de 6pkt/s. De point de vue consommation d'énergie, MEA-DSR était le plus efficace pour tous les taux d'envoi de paquets étudiés. En augmentant le nombre de session, MEA-DSR était le plus performant en termes du DM et du TD tandis que son SRN était le plus élevé. L'efficacité énergétique de MEA-DSR était limitée aux cas de 10 et 15 sessions de communication.

Quant à la distribution de la charge entre les nœuds mobile dans un MANET, nous avons constaté que :

- La mobilité aide à améliorer la diversité des chemins utilisés.
- L'augmentation de la densité offre plus de diversité dans les chemins disponibles.
- Quand le nombre de sessions de communication est important, l'intérêt d'une politique de distribution de la charge devient négligeable.

Enfin, nous notons que l'efficacité de la politique de distribution de la charge implémentée dans MEA-DSR est directement influencée par le nombre des RREQs arrivant aux nœuds destinations. De ce fait, l'augmentation de la congestion et des interférences dans le réseau influent négativement sur cette dernière.

---

[1] C'est à dire le protocole a marqué un TERM plus important, ce qui indique qu'il permettra une durée de vie du réseau plus longue.

**Références**

**Annexe A**

**Le simulateur NS-2**

NS-2 [94] est un simulateur réseaux à événement discrets, open source exécutable sur les plates formes FreeBSD, Linux, Solaris, Mac et Windows via Cygwin (voir fig.A.1). Il permet la simulation des protocoles d'application (Telnet, FTP, etc.), de transport (TCP, UDP), de protocoles de routage pour des réseaux filaires (Distance vector, link state) ou sans fil (DSR, DSDV, TORA, AODV), MAC (IEEE 802.11, TDMA, CSMA, etc.). NS-2 repose sur deux langages : OTCL pour écrire des scripts décrivant la topologie du réseau, et les communications entre les nœuds; et C++ pour l'implantation des différents protocoles et modules.

Comme sortie d'une simulation, NS-2 fournit deux fichiers traces avec les extensions (.tr) et (.nam). Le premier fichier s'utilise dans l'évaluation des performances des protocoles simulés, et cela par des scripts écrits en AWK ou en PERL, tandis que le deuxième est exploité par l'outil "Network Animator" pour la visualisation du réseau simulé.

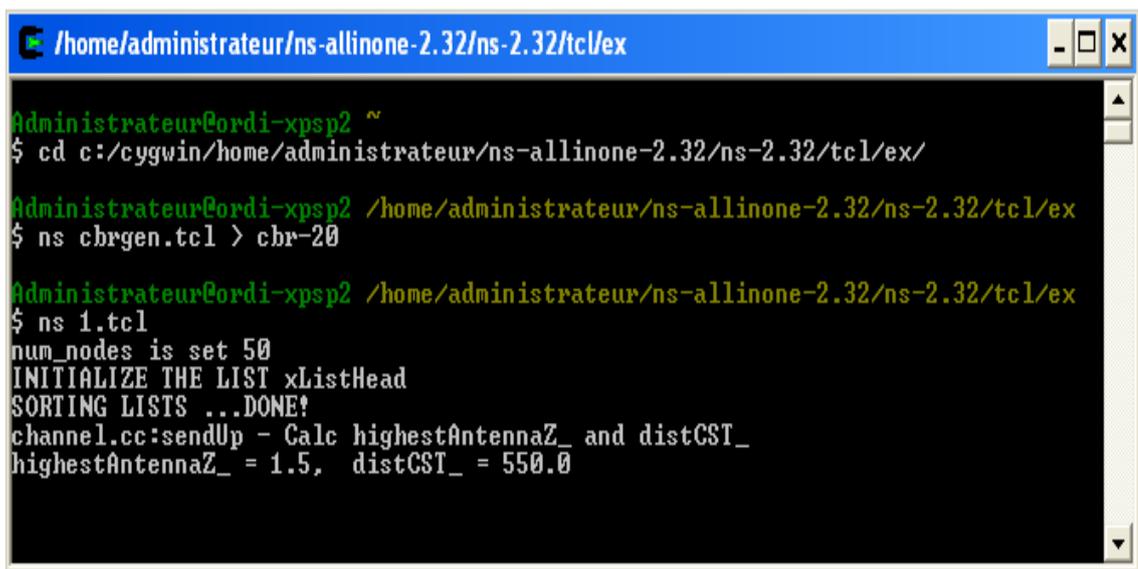

**Figure A.1.** Interface de commande du Cygwin.

**Annexe B**

**Le script qui décrit la configuration du réseau et analyse les résultats de la simulation écrit en OTCL et AWK.**

```
set val(chan)           Channel/WirelessChannel
set val(prop)           Propagation/TwoRayGround
set val(netif)          Phy/WirelessPhy
set val(mac)            Mac/802_11
set val(ifq)            CMUPriQueue
set val(ll)             LL
set val(ant)            Antenna/OmniAntenna
set val(ifqlen)         50
set val(nn)             50
set val(rp)             MEA-DSR
set opt(cp)        "../mobility/scene/ cbr-10"
set opt(sc)        "../mobility/scene/ speed-400-1"
set val(stop)           600.0
set val(x)      1000
set val(y)      1000
set opt(energymodel)    EnergyModel
set opt(initialenergy) 1000
set opt(logenergy)      "on"

Antenna/OmniAntenna set X_ 0
Antenna/OmniAntenna set Y_ 0
Antenna/OmniAntenna set Z_ 1.5
Antenna/OmniAntenna set Gt_ 1.0
Antenna/OmniAntenna set Gr_ 1.0

Phy/WirelessPhy set CPThresh_ 10.0
Phy/WirelessPhy set CSThresh_ 1.559e-11
Phy/WirelessPhy set RXThresh_ 3.652e-10
Phy/WirelessPhy set Rb_ 2*1e6
Phy/WirelessPhy set Pt_ 0.2818
Phy/WirelessPhy set freq_ 914e+6
Phy/WirelessPhy set L_ 1.0

set ns_         [new Simulator]
set tracefd     [open 1.tr w]
$ns_ use-newtrace
$ns_ trace-all $tracefd
set namtrace [open 1.nam w]
$ns_ namtrace-all-wireless $namtrace $val(x) $val(y)
set topo        [new Topography]
$topo load_flatgrid $val(x) $val(y)
create-god $val(nn)
set chan_ [new $val(chan)]
$ns_ node-config -adhocRouting $val(rp) \
            -llType $val(ll) \
-macType $val(mac) \
            -ifqType $val(ifq) \
            -ifqLen $val(ifqlen) \
            -antType $val(ant) \
            -propType $val(prop) \
```

```
            -phyType $val(netif) \
            -topoInstance $topo \
            -energyModel $opt(energymodel) \
            -rxPower 1.0\
            -txPower 1.4\
            -initialEnergy $opt(initialenergy)\
            -agentTrace ON \
            -routerTrace ON \
            -macTrace OFF \
            -movementTrace OFF \
            -channel $chan_

for {set i 0} {$i < $val(nn) } {incr i} {
set node_($i) [$ns_ node]
$node_($i) random-motion 0
}
source $opt(cp)
source $opt(sc)
for {set i 0} {$i < $val(nn)} {incr i} {
$ns_ initial_node_pos $node_($i) 20
}
proc stop { } {
global ns_ tracefd
$ns_ flush-trace
close $tracefd
exec awk {
{
if (($1=="s")&&($19=="AGT")&& ($35=="cbr"))
datas++
if (($1=="r")&&($19=="AGT")&& ($35=="cbr"))
datar++
if (($1 == "s" || $1 == "f") && ($19 == "RTR") && ($35 =="DSR"))
routing_packets++;
if (( $1 == "d" ) && ( $35 == "cbr" )&& ($19 == "RTR")&& ($21 == "TOUT"))
data_dropped_tout++;
if (( $1 == "d" ) && ( $35 == "cbr" )&& ($19 == "RTR")&& ($21 == "TTL"))
data_dropped_ttl++;
if (( $1 == "d" ) && ( $35 == "cbr" )&& ($19 == "RTR")&& ($21 == "NRTE"))
data_dropped_nrte++;
if (( $1 == "d" ) && ( $35 == "cbr" )&& ($19 == "IFQ")&& ($21 == "IFQ"))
data_dropped_ifq++;
if ((start_time[$41]==0)&&( $1 == "s") && ( $35 == "cbr" ) &&
( $19=="AGT" ))
start_time[$41]=$3;
if (( $1 == "r") && ( $35 == "cbr" ) && ( $19=="AGT" ))
end_time[$41] = $3;
if ($1=="N")
cons_energy[$5] =1000-$7;
if ($1=="N")
res_energy[$5] =$7;
}
END{
print "packet delivery fraction", (datar/datas)*100
print "TOUT", data_dropped_tout
print "TTl", data_dropped_ttl
print "NTRE", data_dropped_nrte
print "IFQ", data_dropped_ifq
print "normalized routing overhead",routing_packets/datar
for ( i in end_time )
{
start = start_time[i];
end = end_time[i];
packet_duration = end - start;
```

```
if ( packet_duration > 0 ) { sum += packet_duration; recvnum++; }
}
print "average end to end delay", sum/recvnum
for ( j in cons_energy )
{
total_energy +=cons_energy[j];
}
print "energy consumed per packet", total_energy/datar
moyenne = total_energy/50;
for ( j in cons_energy )
{
sum_ind_dev_square +=(cons_energy[j]-moyenne)*(cons_energy[j]-moyenne);
}
variance= sum_ind_dev_square/50;
print "deviation" , (variance ^ 0.5)
min_res=res_energy[0];
for ( k in res_energy )
{
if (res_energy[k]<min_res )
min_res=res_energy[k];
}
print "minimal residual energy",min_res/10
}
} 1.tr > 1.data
exit 0
}
$ns_ at $val(stop) "stop"
$ns_ run
```

**Annexe C**

**Un script qui décrit différentes sessions de communication écrit en OTCL.**

```
# nodes: 50, max conn: 10, send rate: 0.25, seed: 1.0
# 1 connecting to 2 at time 1.7045592524598163
set udp_(0) [new Agent/UDP]
$ns_ attach-agent $node_(1) $udp_(0)
set null_(0) [new Agent/Null]
$ns_ attach-agent $node_(2) $null_(0)
set cbr_(0) [new Application/Traffic/CBR]
$cbr_(0) set packetSize_ 512
$cbr_(0) set interval_ 0.25
$cbr_(0) set random_ 1
$cbr_(0) set maxpkts_ 10000
$cbr_(0) attach-agent $udp_(0)
$ns_ connect $udp_(0) $null_(0)
$ns_ at 1.7045592524598163 "$cbr_(0) start"
# 4 connecting to 5 at time 37.555412611717088
set udp_(1) [new Agent/UDP]
$ns_ attach-agent $node_(4) $udp_(1)
set null_(1) [new Agent/Null]
$ns_ attach-agent $node_(5) $null_(1)
set cbr_(1) [new Application/Traffic/CBR]
$cbr_(1) set packetSize_ 512
$cbr_(1) set interval_ 0.25
$cbr_(1) set random_ 1
$cbr_(1) set maxpkts_ 10000
$cbr_(1) attach-agent $udp_(1)
$ns_ connect $udp_(1) $null_(1)
$ns_ at 37.555412611717088 "$cbr_(1) start"
# 4 connecting to 6 at time 97.97712619322219
set udp_(2) [new Agent/UDP]
$ns_ attach-agent $node_(4) $udp_(2)
set null_(2) [new Agent/Null]
$ns_ attach-agent $node_(6) $null_(2)
set cbr_(2) [new Application/Traffic/CBR]
$cbr_(2) set packetSize_ 512
$cbr_(2) set interval_ 0.25
$cbr_(2) set random_ 1
$cbr_(2) set maxpkts_ 10000
$cbr_(2) attach-agent $udp_(2)
$ns_ connect $udp_(2) $null_(2)
$ns_ at 97.97712619322219 "$cbr_(2) start"
# 6 connecting to 7 at time 37.089486921713451
set udp_(3) [new Agent/UDP]
$ns_ attach-agent $node_(6) $udp_(3)
set null_(3) [new Agent/Null]
$ns_ attach-agent $node_(7) $null_(3)
set cbr_(3) [new Application/Traffic/CBR]
$cbr_(3) set packetSize_ 512
$cbr_(3) set interval_ 0.25
$cbr_(3) set random_ 1
$cbr_(3) set maxpkts_ 10000
$cbr_(3) attach-agent $udp_(3)
$ns_ connect $udp_(3) $null_(3)

$ns_ at 37.089486921713451 "$cbr_(3) start"
```

```
# 7 connecting to 8 at time 19.697448769443412
set udp_(4) [new Agent/UDP]
$ns_ attach-agent $node_(7) $udp_(4)
set null_(4) [new Agent/Null]
$ns_ attach-agent $node_(8) $null_(4)
set cbr_(4) [new Application/Traffic/CBR]
$cbr_(4) set packetSize_ 512
$cbr_(4) set interval_ 0.25
$cbr_(4) set random_ 1
$cbr_(4) set maxpkts_ 10000
$cbr_(4) attach-agent $udp_(4)
$ns_ connect $udp_(4) $null_(4)
$ns_ at 19.697448769443412 "$cbr_(4) start"
# 7 connecting to 9 at time 5.1353468769860209
set udp_(5) [new Agent/UDP]
$ns_ attach-agent $node_(7) $udp_(5)
set null_(5) [new Agent/Null]
$ns_ attach-agent $node_(9) $null_(5)
set cbr_(5) [new Application/Traffic/CBR]
$cbr_(5) set packetSize_ 512
$cbr_(5) set interval_ 0.25
$cbr_(5) set random_ 1
$cbr_(5) set maxpkts_ 10000
$cbr_(5) attach-agent $udp_(5)
$ns_ connect $udp_(5) $null_(5)
$ns_ at 5.1353468769860209 "$cbr_(5) start"
# 8 connecting to 9 at time 13.656989789408161
set udp_(6) [new Agent/UDP]
$ns_ attach-agent $node_(8) $udp_(6)
set null_(6) [new Agent/Null]
$ns_ attach-agent $node_(9) $null_(6)
set cbr_(6) [new Application/Traffic/CBR]
$cbr_(6) set packetSize_ 512
$cbr_(6) set interval_ 0.25
$cbr_(6) set random_ 1
$cbr_(6) set maxpkts_ 10000
$cbr_(6) attach-agent $udp_(6)
$ns_ connect $udp_(6) $null_(6)
$ns_ at 13.656989789408161 "$cbr_(6) start"
# 9 connecting to 10 at time 50.838808347861658
set udp_(7) [new Agent/UDP]
$ns_ attach-agent $node_(9) $udp_(7)
set null_(7) [new Agent/Null]
$ns_ attach-agent $node_(10) $null_(7)
set cbr_(7) [new Application/Traffic/CBR]
$cbr_(7) set packetSize_ 512
$cbr_(7) set interval_ 0.25
$cbr_(7) set random_ 1
$cbr_(7) set maxpkts_ 10000
$cbr_(7) attach-agent $udp_(7)
$ns_ connect $udp_(7) $null_(7)
$ns_ at 50.838808347861658 "$cbr_(7) start"
# 9 connecting to 11 at time 20.976630459063049
set udp_(8) [new Agent/UDP]
$ns_ attach-agent $node_(9) $udp_(8)
set null_(8) [new Agent/Null]
$ns_ attach-agent $node_(11) $null_(8)
set cbr_(8) [new Application/Traffic/CBR]
$cbr_(8) set packetSize_ 512
$cbr_(8) set interval_ 0.25
$cbr_(8) set random_ 1
$cbr_(8) set maxpkts_ 10000
$cbr_(8) attach-agent $udp_(8)
```

```
$ns_ connect $udp_(8) $null_(8)
$ns_ at 20.976630459063049 "$cbr_(8) start"
# 11 connecting to 12 at time 41.848923043277544
set udp_(9) [new Agent/UDP]
$ns_ attach-agent $node_(11) $udp_(9)
set null_(9) [new Agent/Null]
$ns_ attach-agent $node_(12) $null_(9)
set cbr_(9) [new Application/Traffic/CBR]
$cbr_(9) set packetSize_ 512
$cbr_(9) set interval_ 0.25
$cbr_(9) set random_ 1
$cbr_(9) set maxpkts_ 10000
$cbr_(9) attach-agent $udp_(9)
$ns_ connect $udp_(9) $null_(9)
$ns_ at 41.848923043277544 "$cbr_(9) start"
#Total sources/connections: 7/10
```

**Annexe D**

**Influence du paramètre WT sur les performances du protocole MEA-DSR**

Les courbes montrent que plus WT est important, plus le SRN, l'ECP, l'ETECN et conséquemment le TERM s'améliorent mais cela vient au prix d'une dégradation du TD et d'une augmentation en DM.

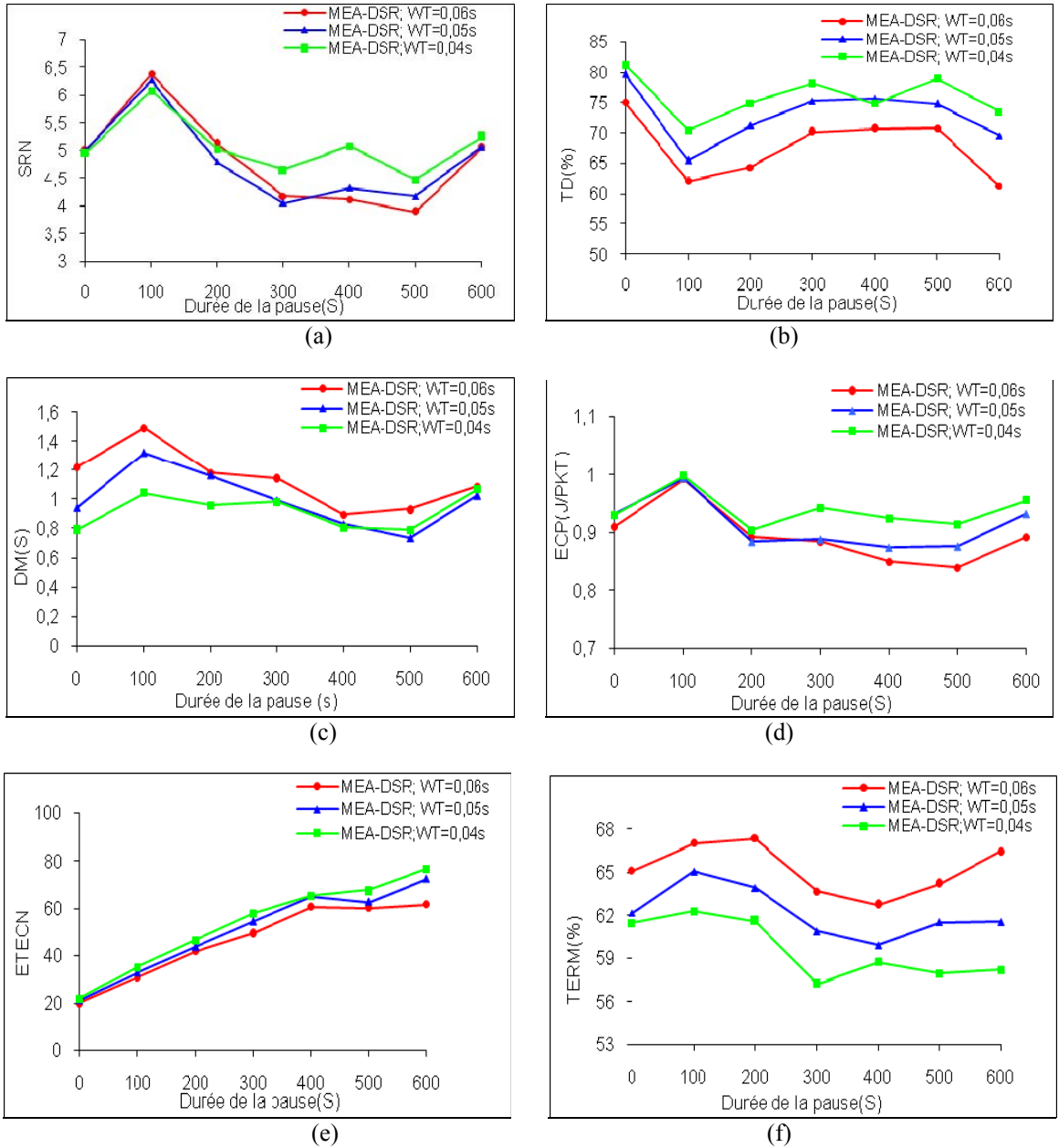

**Figure D.1.** Influence du paramètre WT sur les métriques :
(a) SRN (b) TD (c) DM (d) ECP (e) ETECN (f) TERM.